\documentclass[twocolumn,showpacs,preprintnumbers,amsmath,amssymb]{revtex4}

\usepackage{graphicx}
\usepackage{dcolumn}
\usepackage{bm}
\begin{document}
\tolerance=10000
\hbadness=10000
%
\title{Quadrupole resonances in unstable oxygen isotopes in time-dependent density-matrix formalism}
\author{M. Tohyama}
\affiliation{Kyorin University School of Medicine, 
Mitaka, Tokyo 181-8611, Japan}
\author{A. S. Umar}
\affiliation{Department of Physics and Astronomy, Vanderbilt University,
Nashville, Tennessee 37235, USA}
\date{\today}

\begin{abstract}
The strength functions of quadrupole modes
in the unstable oxygen isotopes $^{22}$O and $^{24}$O are calculated 
using an extended version of the time-dependent Hartree-Fock theory
known as the time-dependent density-matrix theory (TDDM). 
It is found that TDDM gives the lowest quadrupole states which are energetically shifted upward and become 
significantly collective 
due to the coupling to two-body configurations. It is pointed out that
these features of the lowest quadrupole states are similar to
those obtained in the quasi-particle random phase approximation.
\end{abstract}

\maketitle
The study of low-lying quadrupole states in unstable oxygen isotopes
has recently gained both experimental \cite{Jewell,Khan1,Thirolf,Belleguic} 
and theoretical interest \cite{Alex,Utsu,Khan2,Khan3,Matsu}.
One of the reasons for this is that the energy and transition strength of the
first $2^+$ state are 
considered to be closely 
related to the change in the shell structure in this region of neutron rich nuclei \cite{Oza,Otsu1}.
Various theoretical approaches have been used to study the low-lying  
quadrupole states in unstable oxygen isotopes: The shell model \cite{Alex,Utsu},
the random-phase approximation (RPA) \cite{Yoko,Lanza,Hama}, and
the quasi-particle RPA (QRPA) \cite{Khan2,Khan3,Matsu}. It is shown by
the QRPA calculations \cite{Khan2,Khan3,Matsu} that paring correlations 
are essential to reproduce the large $B(E2)$ values of the first $2^+$ states.
Earlier TDDM calculations \cite{Toh1} did not include
the spin-orbit splitting of the single-particle states, thereby
the first $2^+$ states, which mainly
consist of inner shell transitions of neutrons, could not be
treated properly. 
The TDDM approach has recently been improved 
to include the spin-orbit force 
and now is able to handle low-lying quadrupole states in a more quantitative way \cite{Toh2}.
In this letter we apply TDDM to study low-lying quadrupole states, especially, the first $2^+$ states
and show that
TDDM gives results similar to QRPA.
The present formulation of TDDM assumes the Hartree-Fock (HF)
ground state as a starting ground state as will be shown below. 
The quantitative application of TDDM, therefore, is restricted to the quadrupole states in $^{22}$O and $^{24}$O,
where the first-order approximation of the sub-shell closure may be justified \cite{Yoko}.

TDDM is an extended version of 
the time-dependent Hartree-Fock
theory (TDHF)
and is formulated to determine the time evolution of 
one-body and two-body 
density matrices $\rho$ and $\rho_2$ in 
a self-consistent manner \cite{Gon}.
The equations of motion for $\rho$ and $\rho_2$ can be derived
by truncating the well-known Bogoliubov-Born-Green-Kirkwood-Yvon hierarchy for
reduced density matrices \cite{Wan}.
To solve the equations of motion for $\rho$ and $\rho_2$,
$\rho$ and $C_2$ (the correlated part of $\rho_2$) are expanded 
using a finite number of single-particle states $\psi_{\alpha}$
which satisfy a TDHF-like equation,
\begin{eqnarray}
\rho(11',t)=\sum_{\alpha\alpha'}n_{\alpha\alpha'}(t)\psi_{\alpha}(1,t)
\psi_{\alpha'}^{*}(1',t), 
\end{eqnarray}
\begin{eqnarray}
C_{2}(121'2',t)&=&\rho_{2} - {\cal A}(\rho\rho) \nonumber \\
&=&\sum_{\alpha\beta\alpha'\beta'}C_{\alpha\beta\alpha'\beta'}(t)
\nonumber \\
&\times&\psi_{\alpha}(1,t)\psi_{\beta}(2,t)
\psi_{\alpha'}^{*}(1',t)\psi_{\beta'}^{*}(2',t), 
\end{eqnarray}
where ${\cal A}$ is the antisymmetrization operator and the numbers denote space, spin, and isospin coordinates.
Thus, the equations of motion of TDDM consist of the following three
coupled equations \cite{Gon}:
\begin{eqnarray}
i\hbar\frac{\partial}{\partial t}\psi_{\alpha}(1,t)=h(1,t)
\psi_{\alpha}(1,t),
\end{eqnarray}
\begin{eqnarray}
i\hbar \dot{n}_{\alpha\alpha'}=\sum_{\beta\gamma\delta}
[\langle\alpha\beta|v|\gamma\delta\rangle C_{\gamma\delta\alpha'\beta}
-C_{\alpha\beta\gamma\delta}\langle\gamma\delta|v|\alpha'\beta\rangle],
\end{eqnarray}
\begin{eqnarray}
i\hbar\dot{C}_{\alpha\beta\alpha'\beta'}=B_{\alpha\beta\alpha'\beta'}
+P_{\alpha\beta\alpha'\beta'}+H_{\alpha\beta\alpha'\beta'}, 
\end{eqnarray}
where $h$ is the mean-field Hamiltonian and $v$ the residual interaction.
The term $B_{\alpha\beta\alpha'\beta'}$ on the right-hand side of Eq.(5)
represents the Born terms (the first-order terms of $v$). The terms 
$P_{\alpha\beta\alpha'\beta'}$ and $H_{\alpha\beta\alpha'\beta'}$ in Eq.(5)
contain $C_{\alpha\beta\alpha'\beta'}$ and 
represent higher-order particle-particle (and hole-hole) 
and particle-hole type correlations,
respectively. 
Thus full two-body correlations including
those induced by the Pauli exclusion principle
are taken into account in the equation of motion for $C_{\alpha\beta\alpha'\beta'}$.
The explicit expressions for $B_{\alpha\beta\alpha'\beta'}$,
$P_{\alpha\beta\alpha'\beta'}$ and $H_{\alpha\beta\alpha'\beta'}$ are
given in Ref.\cite{Gon}. 
Diverse configurations including excitations of the $^{16}$O core 
can be taken into account in Eqs.(3)-(5).  
In order to obtain a clear understanding of contributions of various configurations 
we solve the coupled equations in the following three schemes: 1)
Two-body amplitudes $C_{\alpha\beta\alpha'\beta'}$
consisting of the neutron 2$s_{1/2}$, 1$d_{3/2}$ and 1$d_{5/2}$ states only
are considered. We call this scheme TDDM1. 2) The neutron 2$p_{3/2}$ and 1$f_{7/2}$ 
are added to TDDM1. We call this version TDDM2. 3) All single-particle states up to the 2$p_{3/2}$ and 1$f_{7/2}$
are taken into account for both protons and neutrons. We call this scheme TDDM3. Two-body configurations corresponding to
excitations of the $^{16}$O core are included in TDDM3.
The number of independent $C_{\alpha\beta\alpha'\beta'}$'s drastically increases from 513 (TDDM1) to
50,727 (TDDM3).

The $E2$ strength function is calculated 
according to the following three
steps 
:

1) A static HF calculation is performed to obtain
the initial ground state. The 1$d_{5/2}$ and 2$s_{1/2}$ orbits are assumed to be
the last fully occupied neutron orbit of $^{22}$O and $^{24}$O, respectively.
The Skyrme III with spin-orbit force
is used as the effective interaction. 
The Skyrme III has been used as one of standard parameterizations of the Skyrme force in nuclear structure calculations 
even for very neutron rich nuclei \cite{Khan2,Otsu}.
The single-particle wavefunctions are confined to a cylinder
with length 20 fm and radius 10 fm. (Axial symmetry is imposed to
calculate the single-particle wavefunctions \cite{Uma}.)
The mesh size used is 0.5 fm.

2) To obtain a correlated ground state, we evolve the HF ground state 
using the TDDM equations and the following time-dependent residual
interaction 
\begin{eqnarray}
v(t)=(1-e^{-t/\tau})v(\bm{r}-\bm{r}').
\end{eqnarray}
The time constant $\tau$ should be sufficiently large 
to obtain a nearly stationary solution of the TDDM equations \cite{Toh4}.
We choose $\tau$ to be 150 fm/$c$. 
In a consistent calculation the residual interaction should be the same as 
that
used to generate the mean field. However, a Skyrme-type force contains momentum dependent terms
, which make the computation time of two-body matrix elements quite large. Therefore,
we need to use a simple force of the $\delta$ function
form $v\propto\delta^3(\bm{r} -\bm{r}')$. 
In order to make a comparison with the results of HFB and QRPA calculations,
we use the following pairing-type residual interaction of the density-dependent $\delta$ 
function form \cite{Chas}
\begin{eqnarray}
v(\bm{r}-\bm{r}')=v_{0}(1-\rho (\bm{r})/\rho_0)
\delta^3(\bm{r}-\bm{r}'),
\end{eqnarray}
where $\rho (\bm{r})$ is the 
nuclear density. $\rho_0$ and $v_{0}$ are set to be 0.16fm$^{-3}$ and $-1200$ MeV fm$^3$,
respectively. Similar values of $\rho_0$ and $v_{0}$ have been used in the Hartree-Fock-Bogoliubov (HFB) 
calculations
\cite{Terasaki,Duguet,Yamagami} in truncated single-particle space.
The time step size used to solve the TDDM equations is
0.75 fm/$c$.

3) The quadrupole mode is excited by boosting the single-particle 
wavefunctions at $t=5\tau$ with the quadrupole velocity field:
\begin{eqnarray}
\psi_{\alpha}(5\tau)\longrightarrow e^{ikQ}\psi_{\alpha}.
\end{eqnarray}
We consider the isoscalar quadrupole mode and the proton quadrupole mode
to calculate a $B(E2)$ value.
For the former, $Q$ is equal to $r^2Y_{20}$ for both protons and neutrons,
and for the latter only the proton single-particle states are boosted.
When the boosting parameter $k$ is sufficiently small,
the strength function defined by
\begin{eqnarray}
S(E)=\sum_{n}|\langle\Phi_n|\hat{Q}|\Phi_0\rangle|^{2}\delta (E-E_{n})
\end{eqnarray}
is obtained from the Fourier transformation of the time-dependent
quadrupole moment $Q(t)$ as
\begin{eqnarray}
S(E)=\frac{1}{\pi k\hbar}\int_{0}^{\infty}Q(t)\sin\frac{Et}{\hbar}
dt. 
\end{eqnarray}
In Eq.(9) $|\Phi_0\rangle$ is the total ground-state wavefunction and
$|\Phi_n\rangle$ is the wavefunction for 
an excited state with excitation energy
$E_n$, and $\hat{Q}$ the quadrupole operator.
The TDDM calculations are stopped at
$t=1500$ fm/$c$ and the upper limit of
the time integration in Eq.(10) becomes 750 fm/$c$.
To reduce fluctuations in $S(E)$, originating from the finite-time integration, 
the quadrupole moment is multiplied by a damping factor 
$e^{-\Gamma t/2\hbar}$
with $\Gamma=1$MeV before the time integration.
Since the integration time is limited, the strength function in a
very low energy region ($E<2\pi\hbar/750\approx2$MeV), which depends on long time behavior
of $Q(t)$, is not well
determined and somewhat affected by numerical inaccuracies.
The energy-weighted sum rule (EWSR) for the isoscalar quadrupole mode is expressed as
\begin{eqnarray}
\int S(E)EdE
&=& \frac{1}{2}\langle\Phi_0|[\hat{Q},[H,\hat{Q}]]|\Phi_0\rangle
\nonumber \\
&=& \frac{5\hbar^2}{4\pi m}A\langle r^2 \rangle,
\end{eqnarray}
where $A$ is the mass number, $m$ is the nucleon mass and
$\langle r^2 \rangle$ means the mean square radius of the ground state.
For the proton mode the mass number is replaced by the atomic number and
the mean square radius is taken for protons, and there appears an enhancement factor due to the
momentum dependence of the Skyrme interaction which is expressed as
\begin{eqnarray}
\frac{5}{8\pi}(t_1+t_2)\int r^2\rho_p(\bm{r})\rho_n(\bm{r})d\bm{r},
\end{eqnarray}
where $t_1$ and $t_2$ are the parameters of the momentum dependent
parts of the Skyrme force, and $\rho_p$ ($\rho_n$ ) is the proton (neutron) density distribution.
In TDDM3 where proton single-particle states are also considered in two-body configurations,
a term containing $C_{\alpha\beta\alpha'\beta'}$, which exhibits the effects of
ground-state correlations appears in addition to Eq.(12) \cite{Toh2}. 
However, its contribution is less than 1\% of the total EWSR value and is neglected.
The contribution of Eq.(12) is about 15\% of the total EWSR value. 
The strength function obtained in TDHF (without Eqs.(4) and (5)) 
is equivalent to that in RPA without any truncation of unoccupied single-particles states
because the TDHF equations for the
boosted single-particle wavefunctions $\psi_{\alpha}$ is solved in coordinate space.
The boundary condition for the continuum states, however, is not properly taken into
account in our calculation because all the single-particle wave functions are confined to 
the cylindrical geometry.

We first discuss some ground state properties of $^{22}$O and $^{24}$O evaluated
at $t=5\tau$ in TDDM1
where only the neutron 1$d_{5/2}$, 2$s_{1/2}$ and 1$d_{3/2}$ orbits are use to evaluate
$C_{\alpha\beta\alpha'\beta'}$.
In $^{22}$O, the energies of the neutron 1$d_{5/2}$, 2$s_{1/2}$, and 1$d_{3/2}$ states are 
$-$6.8MeV, $-$2.3MeV, and 1.3MeV, respectively. Their occupation probabilities are 0.92, 0.14, and 0.05, respectively.
In $^{24}$O, their energies and occupation numbers are 
$-$7.3MeV, $-$2.3MeV, 0.8MeV, 0.97, 0.89, and 0.10, respectively.
The correlation energy $E_{\rm cor}$ defined by
\begin{eqnarray}
E_{\rm cor}=\frac{1}{2}\sum\langle\alpha\beta|v|\gamma\delta\rangle C_{\gamma\delta\alpha\beta}
\end{eqnarray}
is $-4.4$ MeV in $^{22}$O and $-3.5$ MeV in $^{24}$O. 
The correlation energy defined above includes effects of various two-body correlations.
However, pairing correlations seem to dominate. This is because $E_{\rm cor}$ does not change much even if
the sum over single-particle states in Eq.(13) is restricted to a pair of conjugate states under
time reversal such as
\begin{eqnarray}
\frac{1}{2}\sum\langle\alpha\beta|v|\gamma\delta\rangle C_{\gamma\delta\alpha\beta}
\longrightarrow
\frac{1}{2}\sum\langle\alpha\bar{\alpha}|v|\beta\bar{\beta}\rangle C_{\beta\bar{\beta}\alpha\bar{\alpha}},
\end{eqnarray}
where $\bar{\alpha}$ is a state conjugate to $\alpha$ under time reversal.
The correlation energies thus obtained for $^{22}$O and $^{24}$O are $-4.8$ MeV and $-3.9$ MeV,
respectively.
Since the pairing correlations seem dominant, it might be meaningful to 
estimate a quantity $\bar{\Delta}$
corresponding to an average pairing gap in HFB \cite{Duguet,Yamagami}.
We define $\bar{\Delta}$ as 
\begin{eqnarray}
\bar{\Delta}=\frac{\sum\langle\alpha\bar{\alpha}|v|\beta\bar{\beta}\rangle
C_{\beta\bar{\beta}\alpha\bar{\alpha}}}{\sum u_{\alpha}v_{\alpha}},
\end{eqnarray}
where $v_{\alpha}$ and $u_{\alpha}$
are calculated as 
$\sqrt{n_{\alpha\alpha}}$ and $\sqrt{1-n_{\alpha\alpha}}$, respectively.
The values of $\bar{\Delta}$ thus obtained are  $-3.1$ MeV and $-2.7$ MeV for $^{22}$O and $^{24}$O, respectively. 
These values may be comparable to the empirical pairing gap of $\Delta\approx 12/\sqrt{A}\approx 2.5$ MeV.
When the single-particle space is extended to the $2p_{3/2}$ and $1f_{7/2}$ in TDDM2, $\bar{\Delta}$ becomes slightly larger
and its difference between $^{22}$O and $^{24}$O becomes smaller: 
The values of $\bar{\Delta}$ in TDDM2 are $-3.6$ MeV and $-3.5$ MeV for $^{22}$O and $^{24}$O, respectively, and 
become close to the values obtained from HFB calculations using a Woods-Saxon potential \cite{Matsu}.
These values might be rather large as compared to experimental gaps obtained by combining the binding energies
of neighbouring nuclei. However, only a semi-quantitative agreement with experimental values should be expected 
because $\bar{\Delta}$ in Eq.(15) has no direct relation to the definition of the
experimental gap. 
\begin{figure}
    \includegraphics[height=6cm]{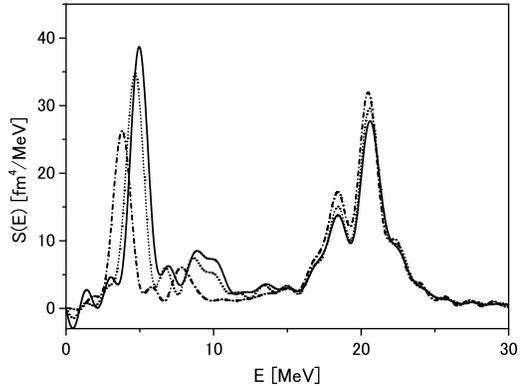}
  \caption{Strength distributions of the isoscalar quadrupole modes
in $^{22}$O
calculated in TDHF (dot-dashed line), TDDM1 (dotted line) and TDDM2 (solid line).}
\end{figure}
\begin{figure}
  \begin{center}
    \includegraphics[height = 6cm]{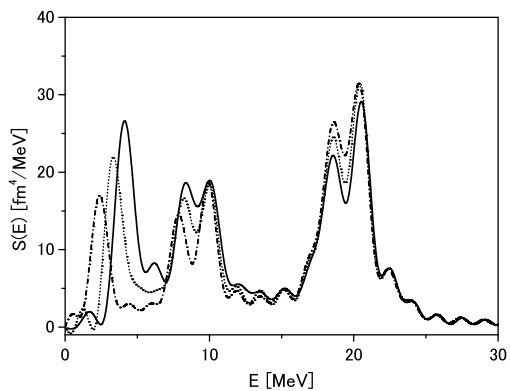}
  \end{center}
  \caption{Strength distributions of the isoscalar quadrupole modes
in $^{24}$O
calculated in TDHF (dot-dashed line), TDDM1 (dotted line) and TDDM2 (solid line).}
\end{figure}

The strength functions for the isoscalar quadrupole modes in $^{22}$O and $^{24}$O 
are shown in Figs.1 and 2, respectively. The dotted lines depict the results in TDDM1 and
the solid lines those in TDDM2. The results in TDHF are also shown with the dot-dashed lines.
The width of each peak is due both to the smoothing with $\Gamma=1$ MeV and to
the finite-time integration of Eq.(10).
In the very low energy region ($E<2$ MeV), $S(E)$ fluctuates slightly and has small negative components.
This originates in uncertainties associated with the calculation of Eq.(10) as mentioned above.
The fraction of the EWSR values depleted below 60MeV is about 98\% in all calculations shown in Figs.1 and 2.
Based on the single-particle energies of the neutron $2s$ and $1d$ states,
we consider that the lowest state seen at $E=3.8$ MeV in TDHF for $^{22}$O 
originates in the transition of a neutron from the 1$d_{5/2}$ orbit to the 2$s_{1/2}$
orbit, while that at $E=2.3$ MeV in $^{24}$O comes from transition of a neutron from the 2$s_{1/2}$ orbit to the 1$d_{3/2}$
orbit. The second lowest states seen at $E=7.8$ MeV in the TDHF calculations for both nuclei are also considered to be
from the inner shell transition of a neutron, that is, the transition from the
1$d_{5/2}$ to the 1$d_{3/2}$. 
When the two-body correlations among the neutron 1$d_{5/2}$, 2$s_{1/2}$ and 1$d_{3/2}$ 
states are included in TDDM1, the collectivity and excitation energies of the lowest states
are increased and the strength of the giant quadrupole resonance (GQR) located at 20 MeV 
is slightly decreased correspondingly. The increase of the excitation energy of the
lowest state is from 3.8 MeV to 4.7 MeV in $^{22}$O and from 2.3 MeV to 3.5 MeV in $^{24}$O.
The increase of the collectivity of the lowest states is due to the mixing of quadrupole states
consisting of two-body configurations 
and the upward shift in excitation energy is related to the lowering of the ground
state due to two-body correlations.
The paring correlations seem to dominate the two-body correlations
because the two-body amplitudes of $C_{\alpha\bar{\alpha}\beta\bar{\beta}}$ type
are most important as discussed above. These behaviors of the first $2^+$ states under
the influence of the pairing correlations are similar to those obtained in QRPA \cite{Khan2,Matsu}.
In TDDM the pairing correlations are treated as two-body correlations, whereas
they are considered in a mean field approximation in QRPA. 
The expansion of the single-particle space up to
the neutron $2p_{3/2}$ and $1f_{7/2}$ states (TDDM2) further enhances the collectivity and the excitation energy:
The excitation energy of the lowest state is increased to 5.0 MeV in $^{22}$O and to 4.4 MeV in $^{24}$O.
The quadrupole strengths below 6 MeV in $^{22}$O are 214 fm$^4$, 271 fm$^4$ and 300 fm$^4$ 
in TDHF, TDDM1 and TDDM2, respectively.
The quadrupole strengths below 6 MeV in $^{24}$O
are 163 fm$^4$, 206 fm$^4$ and 249 fm$^4$ in TDHF, TDDM1 and TDDM2, respectively.  
The increase in the collectivity with increasing number of single-particle states seems to be slightly larger in $^{24}$O
than in $^{22}$O. 

\begin{figure}
  \begin{center}
    \includegraphics[height=6cm]{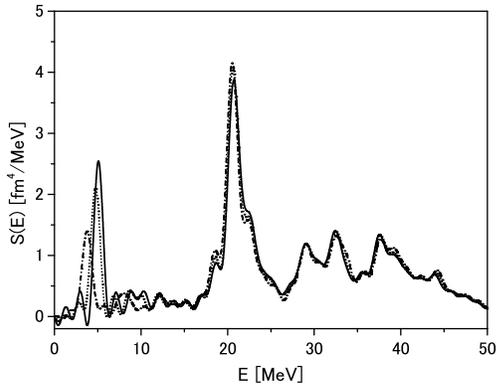}
   \end{center}
  \caption{Strength distributions of the proton quadrupole modes
in $^{22}$O
calculated in TDHF (dot-dashed line), TDDM1 (dotted line) and TDDM2 (solid line).}
\end{figure}
\begin{figure}
  \begin{center}
    \includegraphics[height=6cm]{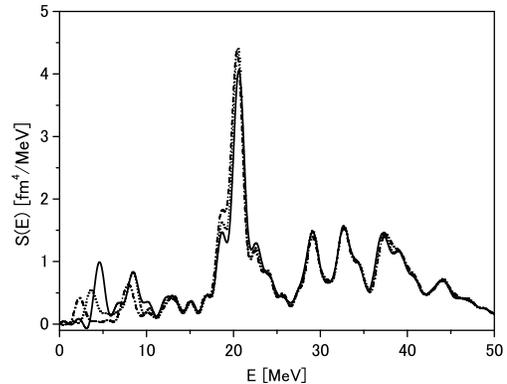}
  \end{center}
  \caption{Strength distributions of the proton quadrupole modes
in $^{24}$O
calculated in TDHF (dot-dashed line), TDDM1 (dotted line) and TDDM2 (solid line).}
\end{figure}

The proton quadrupole modes in $^{22}$O and $^{24}$O are shown in Figs. 3 and 4, respectively.
The fraction of the EWSR values depleted below 60MeV is about 93\% in all calculations shown in Figs. 3 and 4. 
The strength distributions seen above 25 MeV
correspond to isovector quadrupole resonances.
The $B(E2)$ values obtained from the integration of $S(E)$ 
below 6 MeV in $^{22}$O are 11.0 $e^2$fm$^4$, 14.2 $e^2$fm$^4$ and 16.4 $e^2$fm$^4$ 
in TDHF, TDDM1 and TDDM2, respectively.  
The $B(E2)$ value of 16.4 $e^2$fm$^4$ in TDDM2 may be comparable to the observed value 
of $21\pm 8$ $e^2$fm$^4$ \cite{Thirolf} for the first $2^+$ state, although the excitation energy of 5.0 MeV is larger
than the experimental value of 3.2 MeV \cite{Thirolf}. 
The QRPA calculations \cite{Khan2,Matsu} also give the lowest quadrupole states in oxygen isotopes whose
excitation energies are slightly larger than experimental values.
The excitation energies of the first $2^+$ states sensitively depend on the single-particle energies of the $2s$ and $1d$ states.
To reproduce the observed excitation energies, appropriate choice and adjustment of
the Skyrme force parameters including the strength of 
spin-orbit force may be necessary \cite{Khan3}. It has also been pointed out \cite{Colo}
that the energy of the $2s_{1/2}$ state in oxygen isotopes is significantly shifted downward due to
the coupling to phonon states.
The $B(E2)$ values obtained from the integration of $S(E)$ below 6 MeV in  $^{24}$O are 3.5 $e^2$fm$^4$, 5.4 $e^2$fm$^4$ and 7.2 $e^2$fm$^4$ 
in TDHF, TDDM1 and TDDM2, respectively. The small $B(E2)$ value in  $^{24}$O is variant from the QRPA calculation
done by Matsuo \cite{Matsu} but consistent with other QRPA calculations \cite{Khan2,Khan3} and the shell
model calculations \cite{Alex,Utsu}.

\begin{figure}
  \begin{center}
    \includegraphics[height=6cm]{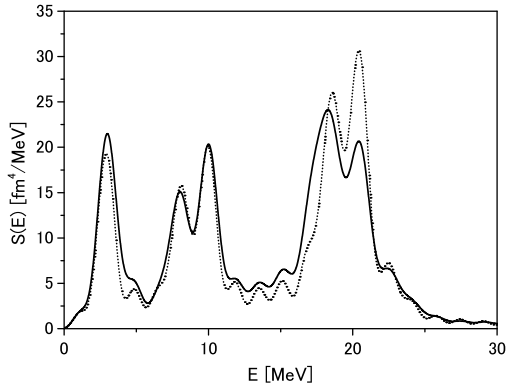}
  \end{center}
  \caption{Strength distributions of the isoscalar quadrupole modes
in $^{24}$O
calculated in TDDM3 (solid line) and TDDM2 (dotted line).
The strength of the residual interaction is $v_0=-800$ fm$^3$MeV.}
\end{figure}

The strength function of the isoscalar quadrupole modes in $^{24}$O calculated in TDDM3 (solid line) 
where the excitations of the $^{16}$O core are included
is shown in Fig.5.
To obtain a stable ground state in TDDM3, we reduced the strength of the residual interaction 
and used $v_0=-800$ fm$^3$MeV in stead of $v_0=-1200$ fm$^3$MeV.
The result in TDDM2 calculated with this value of $v_0$ 
is also shown in Fig.5 with the dotted line. 
The highest peak of GQR is reduced and GQR becomes broader in TDDM3. 
Thus the major effect of the excitations of the $^{16}$O core on the quadrupole mode
is to modify the distribution of the GQR strength.
However, the damping of GQR in $^{24}$O is modest as compared with that in $^{16}$O.
To make this point clear, we show in Fig.6 the
strength function for the isoscalar quadrupole mode in $^{16}$O.
The solid line denotes the result in TDDM3 with $v_0=-800$ fm$^3$MeV
and the dotted line that in TDHF. 
The EWSR value in TDDM3 is 7 \% larger than that in TDHF.
The splitting of the GQR strength obtained in
TDDM3 is consistent with experimental observation \cite{Fritsch}.
The damping of GQR in $^{16}$O is caused by its coupling 
to two-body configurations mainly consisting of two holes in the $1p$ states and two particles 
in the $2s$ or $1d$ states. In $^{24}$O, the coupling of GQR to two-body configurations is 
hindered by the spatial extension
of neutron single-particle wavefunctions due to the neutron excess.

\begin{figure}
  \begin{center}
    \includegraphics[height=6cm]{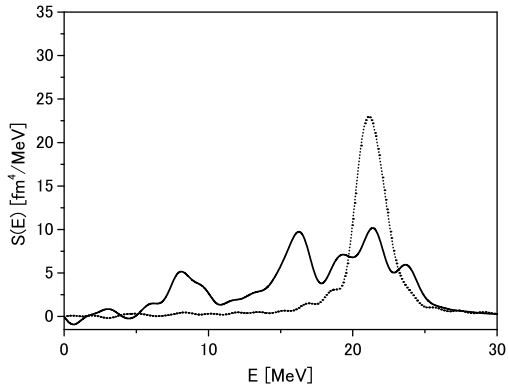}
  \end{center}
  \caption{Strength distributions of the isoscalar quadrupole modes
in $^{16}$O
calculated in TDDM3 (solid line) and TDHF (dotted line).
The strength of the residual interaction used in TDDM3 is $v_0=-800$ fm$^3$MeV.}
\end{figure}

In summary,
the strength functions of the quadrupole resonances
in $^{22}$O and $^{24}$O were studied
using TDDM. In this approach, the correlated ground state was first obtained starting from the HF
ground state with the subsequent excitation of the quadrupole mode. 
It is shown that the lowest quadrupole states are shifted upward and become significantly collective due to the coupling
to two-body configurations. The obtained $B(E2)$ value for $^{22}$O was found comparable to the observed one.
It was pointed out that the lowest quadrupole states obtained in TDDM have properties similar to those in QRPA.
It was also found that the damping of GQR in $^{24}$O is modest as compared with that in $^{16}$O.



\end{document}